\documentclass[preprint,12pt]{elsarticle}
\usepackage{url}
\usepackage{amsmath}
\usepackage{amssymb}
\usepackage{color}
\usepackage{graphicx}
\usepackage{multirow}
\usepackage{threeparttable}
\usepackage{ulem}
\graphicspath{{image/}}
\allowdisplaybreaks

\newcommand{\order}[1]{\mathcal{O}(#1)}

\makeatletter

\renewcommand{\Roman}[1]{\expandafter\@slowromancap\romannumeral #1@}
\makeatother

\journal{Chemical Physics Letter}

\begin{document}

\begin{frontmatter}
\title{A general second order complete active space self-consistent-field solver
for large-scale systems}
\author[caltech]{Qiming Sun}
\author[uhk]{Jun Yang}
\author[caltech]{Garnet Kin-Lic Chan\corref{cor}}
\cortext[cor]{Corresponding author}
\ead{garnetc@caltech.edu}
\address[caltech]{Division of Chemistry and Chemical Engineering, California Institute of Technology,
Pasadena, CA 91125}
\address[uhk]{Department of Chemistry, University of Hong Kong, Pokfulam Road, Hong Kong, China}

\begin{abstract}
We present a new second order complete active space self-consistent field
implementation to converge wavefunctions for both large active spaces and large
atomic orbital (AO) bases. Our algorithm decouples the active space wavefunction
solver from the orbital optimization in the microiterations, and thus may be
easily combined with various modern active space solvers. We also introduce
efficient approximate orbital gradient and Hessian updates, and step size
determination. We demonstrate its capabilities by calculating the low-lying
states of the Fe(\Roman{2})-porphine complex with modest resources using a
density matrix renormalization group solver in a CAS(22,27) active space and a
3000 AO basis.
\end{abstract}


\end{frontmatter}

\section{Introduction} Multiconfigurational electronic
structure is widely found across chemistry\cite{Gordon1998}.
The complete active space self-consistent field
(CASSCF) is a standard tool to describe multiconfigurational electronic structure problems\cite{Gordon1998,Roos1987}.
The CASSCF wavefunction further forms the starting point for more accurate
treatments, including multireference perturbation theory and
configuration interaction methods\cite{Szalay2012}.  Because of its importance, much
 effort has been devoted to efficient CASSCF algorithms in the last 
decades\cite{Lengsfield1980,Werner1980,Roos1980,Olsen1983,Jensen1984,Werner1985,Werner1987,Shepard1987,Nakano2000,Tenno1996,Gyoerffy2013,Lindh2008,Hohenstein2015a,Kim2015}.

A well-known numerical challenge in CASSCF is to converge the self-consistent wavefunction.
For this reason, many early investigations focused on  second order
optimization techniques, which demonstrate superior convergence to pure gradient or super-CI formulations\cite{Lengsfield1980,Werner1980,Roos1980,Jensen1984,Werner1985,Olsen1983,Werner1987,Shepard1987,Nakano2000}.
Unfortunately, these early implementations were optimized for modest AO basis sets,
because they transformed the integrals to the current set of CASSCF orbitals
in each iteration,  incurring significant computational cost and $\order{N^4}$ disk storage.
To extend CASSCF algorithms to large AO bases, several
strategies have been explored\cite{Tenno1996,Gyoerffy2013,Lindh2008,Hohenstein2015a}.
For example, density-fitted
CASSCF\cite{Tenno1996,Gyoerffy2013} and Cholesky decomposition
CASSCF\cite{Lindh2008}  both approximate the AO integrals
to achieve significant savings in the integral
transformation cost and disk storage.
GPU-based AO-driven CASSCF implementations\cite{Hohenstein2015a,Levine} 
further can handle very large numbers of AO functions, although
 these have not yet been extended to second order
optimization. 
Although AO-driven algorithms typically require more floating point operations than
MO-driven approaches, they are favourable for
modern computers, due to their low IO and communication costs.
In this work, our first motivation is to present a new AO-driven algorithm that can
handle large AO basis sets without integral approximations, and also provide second order convergence.
Our algorithm may easily be combined with density-fitting or Cholesky decomposition,
although this is not a focus of this paper.


A second motivation is associated with the need to extend
traditional CASSCF implementations to larger active spaces.
In traditional CASSCF,  full configuration interaction
(FCI) is used as the active space solver.  However, due  to the exponential scaling
of  FCI, it is limited to small
complete active spaces (CAS), usually no more than CAS(16,16) (16 electrons in 16 orbitals).
However, there are now several techniques which can be used to
replace the FCI solver\cite{Chan2011b,Sharma2014,Booth2009,Booth2012a,Vogiatzis2015,DePrince2016}. Two of the more commonly used ones are the
 density matrix renormalization group (DMRG)\cite{Chan2011b} and full configuration interaction quantum
 Monte Carlo (FCIQMC)\cite{Booth2009,Booth2012a}.  These can 
 handle correlated active spaces with many tens of orbitals, and in some cases even more\cite{Sharma2014}.  
While implementations of DMRG and FCIQMC in the CASSCF algorithm exist~\cite{Ghosh2008,Zgid2008,Yanai2009,Ma2013,Wouters2014,fciqmccasscf,Alavi2016,Reiher2016}
they do not yet simultaneously provide second order convergence and the ability to treat very large numbers (i.e. 1000's) of AO's. The implementation we present can be straightforwardly interfaced  to any external active space solver and thus fills this gap.
 In the current work, we will use FCI and DMRG as the active space solvers.
 (An earlier FCIQMC-CASSCF calculation, reported in Ref.~\citenum{fciqmccasscf},  used the two-step version of our implementation
 that we describe here).


In  section \ref{sec:theory}, we describe the formulation of our CASSCF algorithm,
including the approximate orbital gradient and Hessian updates, and orbital optimization
method.
In section \ref{sec:numeric} we
carefully study the convergence
properties and performance of our  algorithm
for several benchmark molecules, within our open-source program package PySCF\cite{PYSCF}.
Finally, as an example of a more challenging large scale problem, we
use FCI and DMRG active space solvers and our CASSCF implementation to converge the
Fe(\Roman{2})-porphine singlet, triplet and quintet ground states.
Our largest calculation uses a 22 electron, 27 orbital active space and almost 3000 AO basis functions.


\section{Algorithm}
\label{sec:theory}

\subsection{Theory}
In this section, we first summarize the relevant formulae for
the optimization of the CASSCF wavefunction.
Given the spin-free electronic Hamiltonian,
\begin{align}
  H &= \sum_{ij}h_{ij}E^i_j + \frac{1}{2}\sum_{ijkl} (ij|kl)(E^i_j E^k_l - \delta_{jk}E^i_l)
  \\
  E^i_j &= a_{i\alpha}^\dagger a_{j\alpha} + a_{i\beta}^\dagger a_{j\beta}
\end{align}
the CASSCF energy can be written as a function of the CI coefficients
$\mathbf{c}$ and the unitary orbital transformation matrix $\mathbf{U}$,
\begin{align}
  E
  &=H_{ijkl}\Gamma_{ijkl}
  \\
  H_{ijkl}
  &=V_{ijkl} U_{pi} U_{qj} U_{rk} U_{sl} \label{eq:fulltrans}
  \\
  V_{pqrs}
  &=\frac{1}{2(N_e-1)} h_{pq} \delta_{rs}
  + \frac{1}{2(N_e-1)} h_{rs} \delta_{pq} + \frac{1}{2}(pq|rs)
  \\
  \Gamma_{ijkl}
  &=\langle I|(E^i_j E^k_l - \delta_{jk}E^i_l)|J\rangle c_Ic_J
\end{align}
where the Einstein summation convention is implied.
Defining a Lagrangian with  normalization constraints for $\mathbf{c}$ and $\mathbf{U}$,
\begin{align}
  F(\mathbf{R},\mathbf{c})
  &= E(\mathbf{R},\Gamma) - \mathcal{E}(\mathbf{c}^\dagger \mathbf{c} - 1)
  \\
  \mathbf{U} &= \exp(\mathbf{R})
  \\
  \mathbf{R} &= -\mathbf{R}^\dagger
\end{align}
 minimizing the energy is a non-linear optimization problem for 
 $\mathbf{R}^*,\mathbf{c}^*$, where the stationary conditions
are
\begin{gather}
  \left.\frac{\partial F}{\partial c_I}
  \right|_{\mathbf{R}^*,\mathbf{c}^*} = 0
  \label{eq:stationary:ci} \\
  \left.\frac{\partial F}{\partial R_{pq}}
  \right|_{\mathbf{R}^*,\mathbf{c}^*} = 0
  \label{eq:stationary:orb}
\end{gather}

The starting point for any second order non-linear optimization algorithm is Newton's method.
Because the energy is quadratic in the CI coefficients, the Newton step for the CI coefficients,
holding the orbitals fixed, is equivalent to solving the standard CI
eigenvalue problem \begin{equation}
  \langle I|(H-\mathcal{E})|J\rangle c_J = 0
  \label{eq:ci}
\end{equation}
Similarly, a Newton step for the orbitals, holding the CI coefficients fixed, corresponds to solving
the equations
\begin{gather}
  \mathcal{H}^{oo} \mathbf{R}^1 + \mathcal{G}^o = 0
  \label{eq:orb:rotation}
  \\
  \mathcal{G}^{o}_{pq}
  = \frac{\partial F}{\partial R_{pq}}
  = \frac{\partial H_{ijkl}}{\partial R_{pq}} \Gamma_{ijkl} \label{eq:orb:gradient}
  \\
  \mathcal{H}^{oo}_{pq,rs}
  = \frac{\partial^2 F}{\partial R_{pq}\partial R_{rs}}
  = \frac{\partial^2 H_{ijkl}}{\partial R_{pq}\partial R_{rs}} \Gamma_{ijkl} \label{eq:orb:hessian}
\end{gather}

The simplest approach to CASSCF optimization is to alternately carry out the Newton steps \eqref{eq:ci}, \eqref{eq:orb:rotation} for the CI coefficients and for the orbitals. 
This simple
alternating scheme is known as the two-step optimization method.
Unfortunately, even when the Newton steps
are carried out exactly, for example, by using the exact orbital Hessian  in Eq. \eqref{eq:orb:rotation},
the two-step method  suffers from slow convergence due to the neglect of coupling between the CI and orbital optimization problems.
It is thus not usually considered a true second order convergent algorithm.

The more sophisticated, one-step, optimization methods aim to approximate
the joint CI and orbital Newton step, corresponding to solving
\begin{equation}
  \begin{pmatrix}
    \mathcal{H}^{cc} & \mathcal{H}^{co} \\
    \mathcal{H}^{oc} & \mathcal{H}^{oo}
  \end{pmatrix}
  \begin{pmatrix}
    \mathbf{c}^1 \\
    \mathbf{R}^1
  \end{pmatrix}
  +
  \begin{pmatrix}
    \mathcal{G}^c \\
    \mathcal{G}^o
  \end{pmatrix}
  = 0
  \label{eq:1step:newton}
\end{equation}
where the Hessian matrices are
\begin{align}
  \mathcal{H}^{cc}_{IJ}
  &=\frac{\partial^2 F}{\partial c_I\partial c_J}
  = \langle I|(H-\mathcal{E})|J\rangle
  \\
  \mathcal{H}^{co}_{I,pq} &= \mathcal{H}^{oc}_{pq,I}
  = \frac{\partial^2 F}{\partial c_I\partial R_{pq}}
  = \frac{\partial H_{ijkl}}{\partial R_{pq}}
  \frac{\partial \Gamma_{ijkl}}{\partial c_I}
\end{align}
Here, the first row of the coupled equations \eqref{eq:1step:newton}
\begin{equation}
  \mathcal{H}^{cc} \mathbf{c}^1 + \mathcal{H}^{co} \mathbf{R}^1 + \mathcal{G}^c = 0
  \label{eq:1step:ci}
\end{equation}
can be rewritten as a CI response problem
\begin{gather}
  \mathbf{H}^0 \mathbf{c}^1 + \mathbf{H}^R \mathbf{c}^0 = E^0\mathbf{c}^1
  \label{eq:ci:response}
\end{gather}
since
{
\begin{gather}
  (\mathcal{H}^{co} \mathbf{R}^1)_I
  = H^R_{ijkl}\langle I|(E^i_j E^k_l - \delta_{jk}E^i_l)|J\rangle c^0_J
  \\
  \mathcal{G}^c_I
  = \langle I|(\mathbf{H}^0-E^0)|J\rangle c_J^0 = 0
\end{gather}
}
where the first order Hamiltonian $\mathbf{H}^R$ is obtained from the
chain rule 
\begin{align}
  H^R_{ijkl}
  &=\frac{\partial H_{ijkl}}{\partial R_{pq}} R^1_{pq}
  = V_{pjkl}R_{pi}^1
  + V_{ipkl}R_{pj}^1
  + V_{ijpl}R_{pk}^1
  + V_{ijkp}R_{pl}^1
  \label{eq:ci:h1}
\end{align}
The second row of Eq. \eqref{eq:1step:newton}
\begin{gather}
  \mathcal{H}^{oo} \mathbf{R}^1 + \mathcal{H}^{oc} \mathbf{c}^1 + \mathcal{G}^o = 0
  \label{eq:1step:orb}
\end{gather}
can be interpreted as the orbital Newton problem with  dressed gradients
\begin{gather}
  \mathcal{H}^{oo} \mathbf{R}^1 = -\tilde{\mathcal{G}}^o
  \label{eq:orb:update}
  \\
  \tilde{\mathcal{G}}^o_{pq}
  = \mathcal{G}^o_{pq} + \mathcal{H}^{oc} \mathbf{c}^1
  = \mathcal{G}^o_{pq} + \frac{\partial H_{ijkl}}{\partial R_{pq}} \Gamma^1_{ijkl}
  \label{eq:dress:g}
  \\
  \Gamma^1_{ijkl}
  = \frac{\partial \Gamma_{ijkl}}{\partial c_I} c^1_I
  \label{eq:orb:gamma1}
\end{gather}
The CI coefficient and orbital optimization problems are thus coupled
through the first order $\mathbf{H}^R$ in Eq. \eqref{eq:ci:response} and the
first order 2-particle density matrix  $\Gamma^1$ in Eq. \eqref{eq:orb:update}.

In principle, in the one-step method, the true CI Newton step requires solving the response equation \eqref{eq:ci:response} exactly.
This is how some early versions of one-step optimization in CASSCF were implemented. 
However, if an iterative procedure is used to determine the CI 
eigenstate in Eq. \eqref{eq:ci},
then a single (or few) steps of the same iterative procedure, with the modified Hamiltonian $\mathbf{H}^0 + \mathbf{H}^R$
and initial eigenstate guess of $\mathbf{c}^0$, can be used to determine an approximate $\mathbf{c}^1$. For example,
a single Davidson iteration\cite{Davidson1975} with these quantities
yields
\begin{align}
\mathbf{c}^1 \approx -[\mathrm{diag}(\mathbf{H}^0-E^0)]^{-1} \mathbf{H}^R \mathbf{c}^0
\label{eq:ci1}
\end{align}
as an approximate solution of Eq. \eqref{eq:ci:response}. The well-known MCSCF implementation by Werner and Knowles\cite{Werner1985},
uses this type of approximation.
In our implementation, we  also use a few iterations of the active space solver to determine an approximate update $\mathbf{c}^1$.
The first order 2-particle density matrix is then computed by finite difference 
\begin{align}
  \Gamma^1_{ijkl} \approx \Gamma_{ijkl}[\mathbf{c}^0 + \mathbf{c}^1]  - \Gamma_{ijkl}[\mathbf{c}^0]
\end{align}
Importantly, this mechanism decouples the orbital optimization from the
active space solver implementation in each Newton step, with the two communicating
solely by passing the 2-particle density matrix and active space Hamiltonian.
This allows us to easily plug-in different iterative active space
solvers, so long as they can provide the 2-particle density matrix. 


A single CI and orbital Newton step provides $\mathbf{c}^1$ and $\mathbf{R}^1$.
We then need to update all quantities that depend on the 
new CI coefficients and new orbitals.
This involves transforming
$H$ to the new set of orbitals  in Eq. \eqref{eq:fulltrans},
solving for the new CI eigenstate  in Eq. \eqref{eq:ci} and computing the new orbital gradients
and Hessians from Eqs. \eqref{eq:orb:gradient} and \eqref{eq:orb:hessian}.
However,  we only perform an exact update of all these quantities every
few Newton steps, as it is computationally very expensive.
Instead for most steps, we use only an approximate update of the quantities.
We call the steps with approximate updates, microiterations.
Every 3 - 4 microiterations, a macroiteration is carried out where all quantities are updated exactly.
We next describe the different kinds of approximate updates  used in the microiterations.



\subsection{Microiterations}
\label{sec:micro}


In each microiteration, we perform approximate updates both for the CI and orbital
parts of the optimization problem. The quality of the update approximation is important, as it can affect
the rate of CASSCF convergence.

We first discuss the orbital  update.
We have considered two frameworks for approximation.
The first is the dynamic-expansion-point (DEP) scheme (as used for example, in Ref. \citenum{Werner1985})
where we compute the 
new Hamiltonian matrix elements $\bar{H}_{ijkl}$ 
\begin{align}
  \bar{H}_{ijkl}
  &=V_{pqrs} \bar{U}_{pi} \bar{U}_{qj} \bar{U}_{rk} \bar{U}_{sl}
  \label{eq:full:transform}
  \\
  \bar{\mathbf{U}} &= \exp(\mathbf{R}^1)
\end{align}
and define the updated orbital gradient and orbital Hessian
from
\begin{align}
  \mathcal{G}^o_{pq}
  &=\left.\frac{\partial F}{\partial R_{pq}}
  \right|_{\mathbf{R}^1}
  = \frac{\partial \bar{H}_{ijkl}}{\partial R_{pq}} \Gamma_{ijkl}
  \\
  \mathcal{H}^{oo}_{pq}
  &=\left.\frac{\partial^2 F}{\partial R_{pq}\partial R_{rs}}
  \right|_{\mathbf{R}^1}
  = \frac{\partial^2 \bar{H}_{ijkl}}{\partial R_{pq}\partial R_{rs}} \Gamma_{ijkl}
  \\
\end{align}
The approximate updates in the DEP framework
consist of approximating  $\bar{H}$ to
reduce the costs of the 4-index integral transformation
\eqref{eq:full:transform}.
By dividing $\bar{\mathbf{U}}$ into two parts
\begin{equation}
  \bar{\mathbf{U}} = 1 + \mathbf{T}
\end{equation}
an approximate $\bar{H}$ can be defined up a given order in
$\mathbf{T}$.
For example, the first order approximate update (DEP1) corresponds to
\begin{equation}
  \bar{H}_{ijkl}
  = V_{ijkl}
  + V_{pjkl} T_{pi}
  + V_{ipkl} T_{pj}
  + V_{ijpl} T_{pk}
  + V_{ijkp} T_{pl}
  \label{eq:dep:h1}
\end{equation}
The exact update is recovered at  fourth order, where the complete 4-index integral transformation
is carried out.
The transformation in DEP4 has the same cost as the integral transformation in the two-step CASSCF optimization method.

The other framework in which to define approximate updates is the fixed expansion point
(FEP) scheme.  In FEP, the orbital gradients and Hessians are defined directly
by an expansion in $\mathbf{R}^1$ 
\begin{gather}
  \mathcal{G}^o \rightarrow \mathcal{G}^o_{pq}
  + \frac{\partial^2 F}{\partial R_{pq}\partial R_{rs}} \cdot R_{rs}^1 + \dots
  \\
  \mathcal{H}^{oo} \rightarrow \mathcal{H}^{oo}_{pq,rs}
  + \frac{\partial^3 F}{\partial R_{pq}\partial R_{rs}\partial R_{tu}} \cdot R_{tu}^1 + \dots
\end{gather}
The approximate updates in the FEP framework correspond to truncating the order of the above expansion.
For example, 
in the simplest FEP1 approximation, we only update the gradients using
the non-updated  Hessian matrix elements
\begin{gather}
  \mathcal{G}^o + \mathcal{H}^{oo} \mathbf{R}^1
  \label{eq:fep0}
\end{gather}
which corresponds to keeping the orbital Hessian frozen within each macro iteration.
We can view the gradients in FEP1 to be an approximation to 
the gradients in DEP1, obtained by replacing $\mathbf{T}$
with $\mathbf{R}$.

To update the CI part, we need to update the Hamiltonian that defines
 the first order CI problem \eqref{eq:ci:h1}.
Analogous approximations within the DEP/FEP framework 
can be formulated for the Hamiltonian update.
For example, Eq. \eqref{eq:dep:h1} is also the definition of the updated CI Hamiltonian within the DEP1 approximation.
%
%
%
%

\subsection{Orbital optimization}
\label{sec:orbopt}

So far we have been discussing the determination of the Newton step.
However, directly following a Newton step is problematic
 in highly non-quadratic optimizations, as the steps can be unbounded, and in fact
are not even guaranteed to go towards the minimum.
This is a well-known problem which is often seen in orbital optimization in CASSCF.

A commonly employed technique to modify the Newton step for the
orbitals, is to use the augmented Hessian (AH) method with step-size control.
Here, a modified Newton step is obtained by solving the eigenvector equation
\begin{equation}
  \begin{pmatrix}
    0 & \mathcal{G}^\dagger \\
    \mathcal{G} & \mathcal{H}
  \end{pmatrix}
  \begin{pmatrix}
    1 \\
    \mathbf{R}^1
  \end{pmatrix}
  = \varepsilon
  \begin{pmatrix}
    1 \\
    \mathbf{R}^1
  \end{pmatrix}
  \label{eq:ah}
\end{equation}
In the standard use of this method, the rotation direction $\mathbf{R}^1$ is first obtained 
as the lowest eigenvector
of Eq. \eqref{eq:ah}.
This direction provides an interpolation between steepest descent and
the full Newton step. Once $\mathbf{R}^1$ is determined, then an approximate line-search is 
performed along $\mathbf{R}^1$ to take an appropriate step size. 

In our current algorithm, we also use the augmented Hessian method,
and we solve Eq. \eqref{eq:ah} using the Davidson method\cite{Davidson1975}.
However, rather than
first determining the orbital search direction $\mathbf{R}^1$ from solving Eq. \eqref{eq:ah},
and then carrying out a separate  line-search, we combine
these two procedures, in a co-iteration.
Each co-iteration corresponds to an AH Davidson iteration, followed
by an update of the orbital gradient in the FEP1 approximation.
For example after $i$ Davidson iterations, the new gradient is updated as
\begin{equation}
  \mathcal{G}^o_{[i+1]} = \mathcal{G}^o_{[i]} + \mathcal{H}^{oo} \mathbf{R}^1_{[i]}
  \label{eq:fep1:update}
\end{equation}
where $\mathbf{R}^{1}_{[i]}$ is the approximate Davidson AH solution at the $i$th iteration.
We ensure that each $\mathbf{R}^{1}_{[i]}$ is sufficiently small, by introducing a scale parameter $\lambda \geq 1$ in the AH equations, where $\lambda$ is chosen such that the largest element in the scaled $\mathbf{R}^{1}_{[i]}$ is smaller
than a predefined threshold (0.03 in our current implementation)
\begin{equation}
  \begin{pmatrix}
    0 & \mathcal{G}_{[i]}^\dagger \\
    \mathcal{G}_{[i]} & \mathcal{H}
  \end{pmatrix}
  \begin{pmatrix}
    1 \\
    \lambda\mathbf{R}^1_{[i]}
  \end{pmatrix}
  = \varepsilon
  \begin{pmatrix}
    1 \\
    \lambda\mathbf{R}^1_{[i]}
  \end{pmatrix}
\end{equation}
Accumulating the small steps $\mathbf{R}_{[i]}^1$ from the co-iterations, we obtain the full orbital rotation,
\begin{equation}
  \mathbf{R}^1 = \mathbf{R}^1_{[0]} + \mathbf{R}^1_{[1]} + \mathbf{R}^1_{[2]} + \dots
  \label{eq:racc}
\end{equation}
Because the gradients are updated, the quantities entering into the AH equations change at each co-iteration.
Table \ref{tab:fwah} shows how $\mathcal{H}$, $\mathcal{G}$ and $\mathbf{R}_{[i]}^1$ as a function of the iteration number.


We choose $\lambda$ such that every step $\mathbf{R}_{[i]}^1$ is small, but
in our numerical tests, this small stepsize 
does not lead to low efficiency.
Instead, the accumulated small steps effectively provide the ability to take large total rotations in the
orbital space. We thus achieve a good compromise between
larger steps for greater efficiency, and small steps for greater robustness.

\subsection{Computational costs}
\label{sec:cost}

The two main  computational costs to consider are the
memory usage and the operation count.
Memory usage is quite different in MO-driven and AO-driven CASSCF 
optimization algorithms.
In the traditional MO-driven algorithm,  all integrals are transformed to the MO
representation in each macroiteration, and then they are reused in the microiterations.
Although the MO-driven approach requires less CPU resources,  it requires more I/O than the AO-driven
algorithm when there are a large number of core orbitals.
For example, to evaluate the contraction $\mathcal{H}\mathbf{R}^1$ in the DEP1
approximation, the MO-driven algorithm requires the two-electron integrals
$(AA|\!*\!*)$, $(A\!*|A*)$, $(CC|VV)$, $(CV|CV)$, $(CV|AC)$, $(CV|AV)$,
$(AV|CC)$, $(AC|VV)$
(see the matrix elements in the supplemental material),
where the letters $C,A,V$ stand for core, active and external orbitals, and the symbol
``$*$'' stands for any of these kinds of orbitals.
However, the integrals $(CC|VV)$, $(CV|CV)$, $(CV|AC)$, $(CV|AV)$, $(AV|CC)$,
$(AC|VV)$ are associated with the contractions to the core or the active space
1-particle density matrices.
These can instead be evaluated in a direct-SCF style AO-driven J and K (Coulomb and
exchange matrix) build.
As such, the AO-driven algorithm only requires two kinds of integrals
$(AA|\!*\!*)$, $(A\!*|A*)$ to be computed in each macroiteration.
For the various orders of the DEP/FEP approximations, Table \ref{tab:costs} summarizes the
memory requirements in the MO-driven and AO-driven implementations.
Due to the lower memory requirements, the AO-driven DEP1 approximation is favoured
in our general CASSCF implementation.

The CPU costs of the CASSCF algorithm are more difficult to optimize than
the memory costs.
In the one-step algorithm, the
demanding CPU steps are the expensive macro iteration (which invoke the 4-index
integral transformation and the accurate solution for the  CI eigenstate),
the CI response problem in the micro iteration,
and the contraction operation
$\mathcal{H}^{oo}\mathbf{R}^1$ in the orbital optimization problem.
The CI response problem in the micro iteration can be a dominant cost
when an expensive active space solver, such as DMRG or FCIQMC
is used. 
However, as we have mentioned above, we can solve the CI response equation
approximately, by solving the CI eigenstate problem approximately.
Using a FCI active space solver, 2 - 5 Davidson iterations are usually enough to
generate a sufficiently accurate first order density matrix to generate the dressed orbital gradients
\eqref{eq:dress:g}, while using a DMRG active space solver, 2 - 3 sweeps
are usually sufficient.


The $\mathcal{H}^{oo}\mathbf{R}^1$ contraction in the AO-driven algorithm is
expensive because solving the AH equations in the orbital optimization requires many
$\mathcal{H}^{oo}\mathbf{R}^1$ contractions, and every
$\mathcal{H}^{oo}\mathbf{R}^1$ contraction requires a build of the
entire J and K matrices.
One workaround is to gradually increase the accuracy of the solution of the AH problem in
the optimization.
The AH equation is first only approximately solved in a small Davidson subspace, until
the CASSCF wavefunction is close to convergence.

In our implementation, we carry out 4 microiterations for each macroiteration and 4 - 6 Davidson (co)-iterations per microiteration.
Thus, the ratio of the number of macroiterations, microiterations and J and K builds, is
roughly 1:4:20. It should be noted that our parameters have not been carefully optimized
to minimize the total number of macroiterations.
Instead, we have used  conservative settings to provide robust convergence. 
Optimal settings will be investigated in future studies.

\section{Numerical studies}
\label{sec:numeric}
We  next assess the performance of our CASSCF algorithm
in a variety of small molecules and challenging systems.
We first tested the convergence properties as a function of the
approximate orbital gradient and Hessian DEP/FEP updates.
The computational efficiency was measured by the number
of macroiterations, microiterations, and J and K builds.
Next, to test the capabilities of the algorithm
in a large problem, we optimized the singlet, triplet, and quintet
states of the Fe(\Roman{2})-porphine molecule in a (8e, 11o) active space
using the cc-pVDZ (439 AO functions),
cc-pVTZ (956 AO functions),  cc-pVQZ (1784 AO functions),
and cc-pV5Z bases (2997 AO functions).
Finally, we demonstrate the flexibility of our implementation
by carrying out a larger active space study (22e, 27o) using
a DMRG active space solver.


The general CASSCF optimization algorithm and FCI solver were implemented in the
open-source PySCF program package\cite{PYSCF}.
The DMRG solver was provided by the \textsc{Block} code\cite{Sharma2012}.  All tests
were executed on a workstation  equipped with 2 Xeon E5-2670 CPUs
(16 CPU cores @ 2.5 GHz) and 64 GB memory.

\subsection{DEP and FEP approximations}

We first summarize the different
 DEP/FEP approximate updates that we tested in  Table \ref{tab:approx1step}.
At the lowest level of FEP1, we only consider the contribution of the first order
$\mathbf{R}^1$ to the orbital gradients and CI Hamiltonian.
In the DEP1 and DEP2 approximations, the orbital gradients and CI Hamiltonian are
evaluated up to  first and second order in $\mathbf{T}$, respectively.
The contributions of $\mathbf{T}$ to the orbital Hessian are not included in the
FEP1, DEP1, and DEP2 approximations.
The treatment of the orbital gradients and CI Hamiltonians in DEP1+, DEP2+ is
the same as in the DEP1, DEP2 approximations, while higher order terms in the orbital
Hessian are included. In DEP4+, all quantities are exactly updated, because 4th order is
the highest expansion order for the Hamiltonian matrix elements. 


Table \ref{tab:dep1test} presents the convergence performance of the 1-step optimization method,
for different DEP and FEP approximations, in a variety of simple systems.
Regardless of the approximation, the 1-step method
is always superior to the 2-step method.
The number of macroiterations required in the 1-step method is only 1/3 to 1/4 of that
required by the 2-step method.
Except for the two systems CH$_2$ and O$_3$, the different approximations present similar
comparative convergence behavior across the different molecules.
For CH$_2$ and O$_3$, the lower order approximations FEP1 and DEP1 behave
slightly worse than DEP2 and DEP4.

\begin{figure}[htp]
  \begin{center}
  \includegraphics[width=\textwidth]{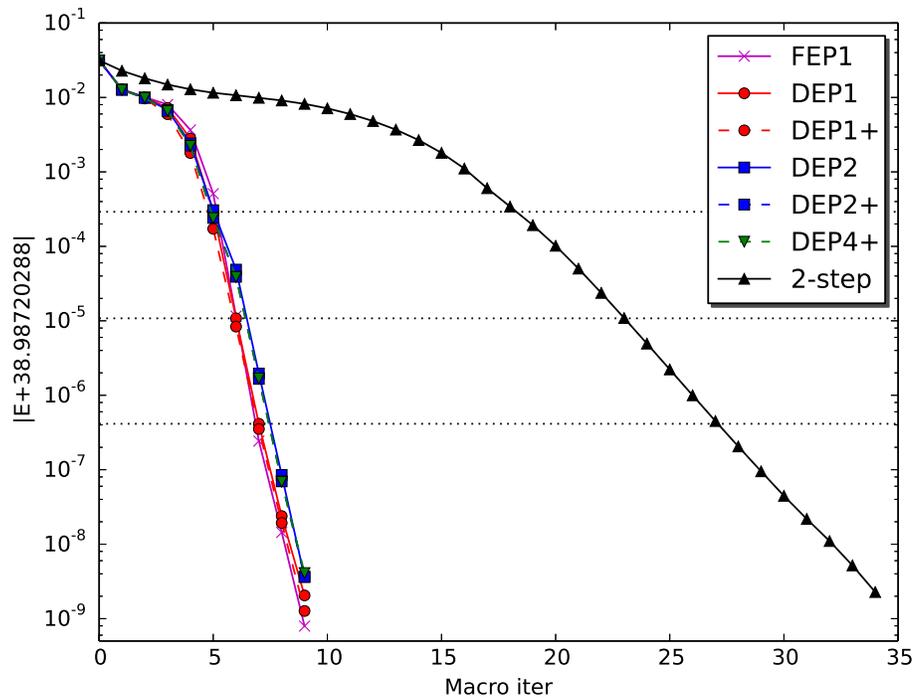}
  \end{center}
  \caption{Convergence of the energy of the  $^1A_1$ state of CH$_2$ molecule with CAS(6,14)/cc-pVDZ.}
  \label{fig:conv:ch2}
\end{figure}

Figure \ref{fig:conv:ch2} visualizes the convergence of the total energy
against the macroiteration number for CH$_2$ CAS(6,14)/cc-pVDZ.
The six different approximate updates used in the microiterations all show similar
  convergence properties, and they all require 9 macroiterations to converge the total energy.
One microiteration of the 1-step method shows similar convergence gains to one macroiteration of the 2-step
method, despite being much cheaper computationally.
Since we used 4 microiterations for every macroiteration in the 1-step method, each
 macroiteration of the 1-step method effectively performs like 4 macroiterations in the
2-step method.  As shown by the horizontal dotted lines in
the figure, the DEP1 approximation needed 2 macroiterations in the 1-step method to converge the
energy error from $10^{-3}$ to $10^{-6}$ Hartrees, while the 2-step method using the same DEP1 approximation, took 9
macroiterations to achieve the same.

\begin{figure}[htp]
  \begin{center}
  \includegraphics[width=\textwidth]{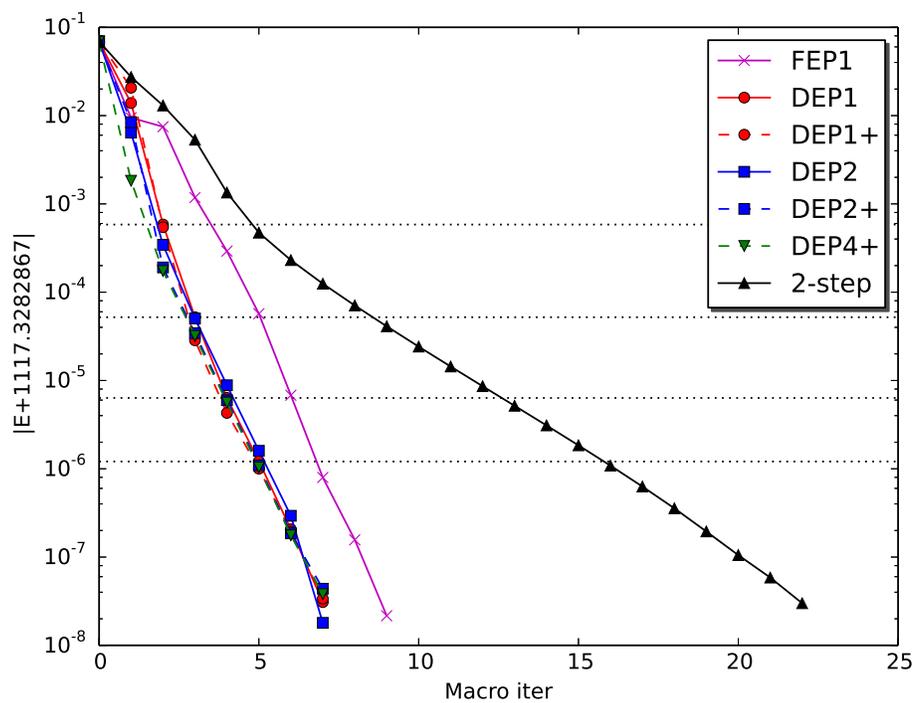}
  \end{center}
  \caption{Convergence of the ground-state energy of Al$_4$O$_2$ with CAS(12,12)/aug-cc-pVDZ.}
  \label{fig:conv:al4o2}
\end{figure}

Figure \ref{fig:conv:al4o2} shows the convergence behaviour for the 
singlet ground state of the Al$_4$O$_2$ cluster (see supplemental material for
the geometry) using a CAS(12,12) active space and the  aug-cc-pVDZ basis.
The active space is chosen using a DMET-like procedure\cite{Knizia2012b,QimingSun2014,practicalDMET,Sayfutyarova2017}.
First, the AO's were localized with a meta-L\"owdin orthogonalization\cite{QimingSun2014},
then the 6 oxygen 2$p$ orbitals were selected as the ``impurities''.
The resulting 6 impurity and 6 bath orbitals obtained as the DMET space were
chosen to be the initial active space.
Except when using the FEP1 approximate update, the 1-step optimization converges in 7 macroiterations, about 1/3 
of the 22 macroiterations required by the 2-step method.
Using the DEP1 approximation, converging the energy error from $10^{-3}$ to
$10^{-6}$ Hartrees required 3 macroiterations of the 1-step method, and 11 macroiterations of the 2-step method.

\begin{figure}[htp]
  \begin{center}
  \includegraphics[width=\textwidth]{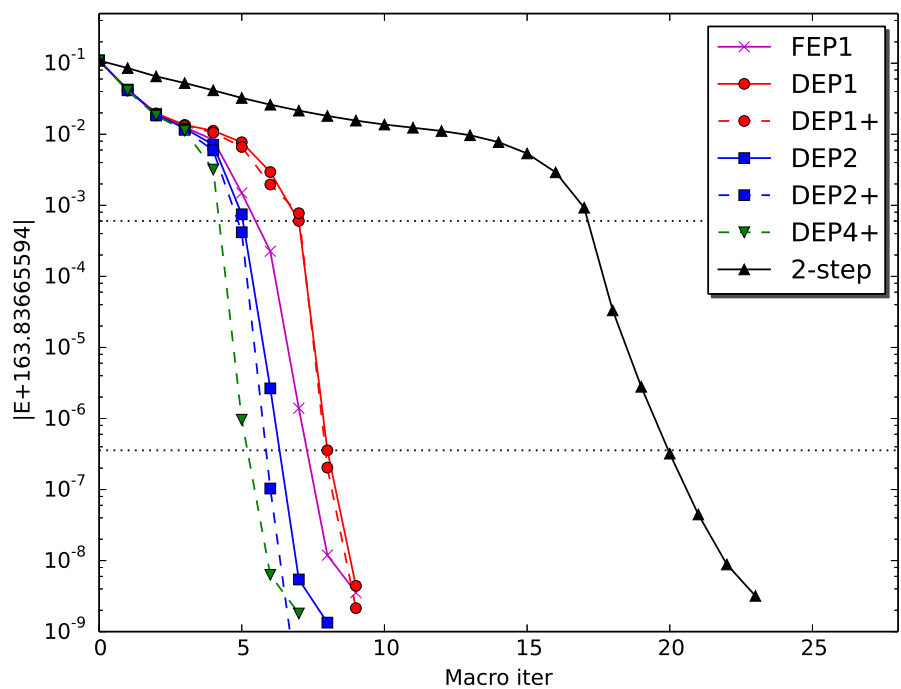}
  \end{center}
  \caption{Convergence of the ground-state energy of HN$_3$ with CAS(10,10)/cc-pVTZ.}
  \label{fig:conv:hn3}
\end{figure}

Figure \ref{fig:conv:hn3} show the convergence behaviour of the optimization algorithm for the
ground-state ($^1A_1$) of the HN$_3$ molecule (see supplemental material for
the geometry) using a CAS(10,10) active space and the cc-pVTZ basis.
We use 10 Hartree-Fock orbitals (5 HOMO's, 5 LUMO's) to form the initial active space. We observed a large
 change in the character of the active space and slow initial convergence for
both the 1-step and 2-step methods.
Depending on the approximate update used in the microiteration, the 1-step method needed
7 - 9 macroiterations to converge.

Overall, our tests demonstrate that the  low level
approximate update (DEP1) performs almost as well as the exact orbital update (DEP4+)
with our current CASSCF algorithm settings.
However, it should be noted that the total number of macroiterations
has not been optimized for the case of the high-level DEP approximations.
For example, it is always possible to converge DEP4+ in one macro
iteration, with a very large number of microiterations, because all
quantities are updated exactly in the microiterations.
However, we cannot use too many microiterations with DEP1 because the errors in
the approximate updates will quickly accumulate.
To obtain the best runtime performance, a trade-off has to be made between the
DEP level (for memory/disk usage) and the number of macroiterations (for CPU
time). In the next section, we use the DEP1 approximation because of
its low memory requirements, which allows us to treat a large number of orbitals.

\subsection{CASSCF and DMRG-CASSCF study of Fe(\Roman{2})-porphine}
\label{sec:performance}
The Fe(\Roman{2})-porphine ground
state\cite{Edwards1986,Rovira1997,Choe1999,Oliveira2001,Pierloot2003,Groenhof2005,Liao2006,Shaik2011,Swart2013,Olivares-Amaya2015} has long been a target of multireference quantum chemistry studies. There has been much debate
about the ordering of the lowest spin states.
Density functional approximations tend to predict a triplet ground state, while
many CASSCF and multireference perturbation calculation with small active
spaces argue for a quintet ground-state\cite{Choe1999,Pierloot2003,Groenhof2005,Vancoillie2011}.

We carried out CASSCF calculations at the triplet geometry\cite{Groenhof2005}
with D$_{2h}$ symmetry in the cc-pVDZ, cc-pVTZ, cc-pVQZ, and cc-pV5Z bases.
The active space was initialized with the DMET procedure and consisted of Fe 3$d$ and
4$d$ meta-L\"owdin orthogonalized orbitals and one optimal bath orbital\cite{QimingSun2014}, giving 11 orbitals in total.
The DMET bath orbital was chosen as the most
important bath orbital from the spin-free ROHF density matrix,
which is strongly
entangled with the Fe 3$d$ and 4$d$ orbitals.
Based on the Hartree-Fock density matrix, we assigned 8 electrons to the
active space. The population density was 6.14 electrons on the impurity (Fe 3d, 4d) orbitals, and 1.88 on
the bath orbital.
Table \ref{tab:feporcas11} gives the energies of some of the low lying singlet, triplet, and
quintet states.
With our initial guess, good convergence was found for all states except for the
$^3B_{1g}$ state.
For the $^3B_{1g}$ state, within 4 - 5 iterations, the optimization appeared to approach
a very flat region, with small gradients and energy changes of about $10^{-6}$ Hartrees
between  macroiterations.
However, after 5 - 10 more macroiterations, the optimization left this region,
and then rapidly converged to a solution that was  about 1 m$E_\mathrm{h}$ lower than
the flat region. The converged solution has significant 4$s$ components, which are not part of the initial guess.

Finally, as an example of a CASSCF calculation with a larger active space,
we extended our active space to  27 orbitals, containing the Fe $3d,4d,4s$ shells (11 orbitals),
4 N $2p_z$ orbitals, N $2p_x$ and $2p_y$ orbitals (for the Fe-N $\sigma$ bonds)
and the most important 8 DMET bath orbitals generated using the above impurity orbitals.
The 8 DMET bath orbitals constituted 4  ligand $\pi$ orbitals and 4 Fe-N anti-bonding ligand orbitals.
 To tackle this large active space, we used a DMRG active space solver, with a bond dimension of $M=1000$.
For the AO basis, we used the cc-pVDZ, cc-pVQZ, and cc-pV5Z bases.
The energies of the triplet and quintet ground-states are given in Table \ref{tab:feporcas11}.
In this larger active space, we find that  $^5A_{g}$ is the lowest state and is about 6 m$E_\mathrm{h}$ lower than $^3B_{3g}$ state, 
irrespective of basis.

In Table \ref{tab:feporcas11}, we show the efficiency of the CASSCF optimization in these Fe(\Roman{2})-porphine 
calculations, as measured by the number of
J and K matrix evaluations, microiterations, and macroiterations. 
Depending on the type of calculation, the time-dominant step differs.
For small active spaces, the CPU time for the CI problem is negligible.
The J and K matrix evaluation is also more costly than the integral transformation.
For example, it takes 49 hours to converge the $^5A_{g}$ state for
CAS(8,11)/cc-pVQZ on a 16-core node, in which about 63.7 \% of the time is
used to evaluate J and K matrices, and 35 \% of the time is spent on the macroiterations (in integral transformation).
When the active space is larger and handled by a more expensive active space solver,
the cost of the macroiterations and microiterations both increase.
In the DMRG-CAS(22,27) calculation, 
 41 \% of the time was spent on macroiterations (with about 1/5 of the time in
the DMRG solver and 4/5 of the time in integral transformation). In the microiterations, over 8 \% of the time 
was spent in the DMRG solver.




\section{Conclusion}

In this work, we presented a general  second order CASSCF implementation for large scale calculations.
We used an AO-driven approach to handle large AO bases, and formulated our Newton steps
to decouple the CI solver from the orbital optimization in each microiteration,
thus allowing ready interfacing to modern active space solvers, such as the DMRG and FCIQMC.
Further, to achieve greater efficiency, 
we proposed several approximate updates of the orbital and CI gradient and Hessians,
as well as a co-iterative augmented Hessian algorithm to determine
the orbital step.
We assessed the numerical performance of the general CASSCF solver with different approximate updates,
and with 1-step and 2-step optimization,
in a variety of small molecules, and in a larger case-study of the
Fe(\Roman{2})-porphine low-lying states. Using our algorithm, we showed that we could
converge a DMRG-CASSCF calculation using a (22,27) active space and 3000 AO basis functions
with only modest resources. 
\clearpage

\section{Acknowledgments}

This work was supported by the National Science Foundation through NSF-CHE-1657286. Additional support
was provided by NSF-CHE-1650436. Further
support for GKC was provided by the Simons Foundation through a Simons Investigatorship.

\bibliographystyle{elsarticle-num}
\bibliography{ref}

\begin{thebibliography}{10}
\expandafter\ifx\csname url\endcsname\relax
  \def\url#1{\texttt{#1}}\fi
\expandafter\ifx\csname urlprefix\endcsname\relax\def\urlprefix{URL }\fi
\expandafter\ifx\csname href\endcsname\relax
  \def\href#1#2{#2} \def\path#1{#1}\fi

\bibitem{Gordon1998}
M.~W. Schmidt, M.~S. Gordon,
  \href{http://dx.doi.org/10.1146/annurev.physchem.49.1.233}{The construction
  and interpretation of mcscf wavefunctions}, Annu. Rev. Phys. Chem. 49~(1)
  (1998) 233--266.
\newblock \href
  {http://arxiv.org/abs/http://dx.doi.org/10.1146/annurev.physchem.49.1.233}
  {\path{arXiv:http://dx.doi.org/10.1146/annurev.physchem.49.1.233}}, \href
  {http://dx.doi.org/10.1146/annurev.physchem.49.1.233}
  {\path{doi:10.1146/annurev.physchem.49.1.233}}.
\newline\urlprefix\url{http://dx.doi.org/10.1146/annurev.physchem.49.1.233}

\bibitem{Roos1987}
B.~O. Roos, \href{http://dx.doi.org/10.1002/9780470142943.ch7}{The complete
  active space self-consistent field method and its applications in electronic
  structure calculations}, Adv. Chem. Phys. 69 (1987) 399--445.
\newblock \href {http://dx.doi.org/10.1002/9780470142943.ch7}
  {\path{doi:10.1002/9780470142943.ch7}}.
\newline\urlprefix\url{http://dx.doi.org/10.1002/9780470142943.ch7}

\bibitem{Szalay2012}
P.~G. Szalay, T.~M\"uller, G.~Gidofalvi, H.~Lischka, R.~Shepard,
  \href{http://pubs.acs.org/doi/abs/10.1021/cr200137a}{Multiconfiguration
  self-consistent field and multireference configuration interaction methods
  and applications}, Chem. Rev. 112~(1) (2012) 108--181.
\newblock \href
  {http://arxiv.org/abs/http://pubs.acs.org/doi/pdf/10.1021/cr200137a}
  {\path{arXiv:http://pubs.acs.org/doi/pdf/10.1021/cr200137a}}, \href
  {http://dx.doi.org/10.1021/cr200137a} {\path{doi:10.1021/cr200137a}}.
\newline\urlprefix\url{http://pubs.acs.org/doi/abs/10.1021/cr200137a}

\bibitem{Lengsfield1980}
B.~H. Lengsfield, \href{http://link.aip.org/link/?JCP/73/382/1}{General second
  order mcscf theory: A density matrix directed algorithm}, J. Chem. Phys.
  73~(1) (1980) 382--390.
\newblock \href {http://dx.doi.org/10.1063/1.439885}
  {\path{doi:10.1063/1.439885}}.
\newline\urlprefix\url{http://link.aip.org/link/?JCP/73/382/1}

\bibitem{Werner1980}
H.-J. Werner, W.~Meyer,
  \href{http://scitation.aip.org/content/aip/journal/jcp/73/5/10.1063/1.440384}{A
  quadratically convergent multiconfiguration-self-consistent field method with
  simultaneous optimization of orbitals and ci coefficients}, J. Chem. Phys.
  73~(5) (1980) 2342--2356.
\newblock \href {http://dx.doi.org/http://dx.doi.org/10.1063/1.440384}
  {\path{doi:http://dx.doi.org/10.1063/1.440384}}.
\newline\urlprefix\url{http://scitation.aip.org/content/aip/journal/jcp/73/5/10.1063/1.440384}

\bibitem{Roos1980}
B.~O. Roos, P.~R. Taylor, P.~E. Siegbahn,
  \href{http://www.sciencedirect.com/science/article/pii/0301010480800450}{A
  complete active space \{SCF\} method (casscf) using a density matrix
  formulated super-ci approach}, Chem. Phys. 48~(2) (1980) 157 -- 173.
\newblock \href
  {http://dx.doi.org/http://dx.doi.org/10.1016/0301-0104(80)80045-0}
  {\path{doi:http://dx.doi.org/10.1016/0301-0104(80)80045-0}}.
\newline\urlprefix\url{http://www.sciencedirect.com/science/article/pii/0301010480800450}

\bibitem{Olsen1983}
J.~Olsen, D.~L. Yeager, P.~J{\o}rgensen,
  \href{http://dx.doi.org/10.1002/9780470142783.ch1}{Optimization and
  characterization of a multiconfigurational self-consistent field (mcscf)
  state}, Adv. Chem. Phys. 54 (1983) 1--176.
\newblock \href {http://dx.doi.org/10.1002/9780470142783.ch1}
  {\path{doi:10.1002/9780470142783.ch1}}.
\newline\urlprefix\url{http://dx.doi.org/10.1002/9780470142783.ch1}

\bibitem{Jensen1984}
H.-J.~A. Jensen, P.~J{\o}rgensen,
  \href{http://scitation.aip.org/content/aip/journal/jcp/80/3/10.1063/1.446797}{A
  direct approach to second-order mcscf calculations using a norm extended
  optimization scheme}, J. Chem. Phys. 80~(3) (1984) 1204--1214.
\newblock \href {http://dx.doi.org/http://dx.doi.org/10.1063/1.446797}
  {\path{doi:http://dx.doi.org/10.1063/1.446797}}.
\newline\urlprefix\url{http://scitation.aip.org/content/aip/journal/jcp/80/3/10.1063/1.446797}

\bibitem{Werner1985}
H.-J. Werner, P.~J. Knowles,
  \href{http://scitation.aip.org/content/aip/journal/jcp/82/11/10.1063/1.448627}{A
  second order multiconfiguration scf procedure with optimum convergence}, J.
  Chem. Phys. 82~(11) (1985) 5053--5063.
\newblock \href {http://dx.doi.org/http://dx.doi.org/10.1063/1.448627}
  {\path{doi:http://dx.doi.org/10.1063/1.448627}}.
\newline\urlprefix\url{http://scitation.aip.org/content/aip/journal/jcp/82/11/10.1063/1.448627}

\bibitem{Werner1987}
H.-J. Werner,
  \href{http://dx.doi.org/10.1002/9780470142943.ch1}{Matrix-formulated direct
  multiconfiguration self-consistent field and multiconfiguration reference
  configuration-interaction methods}, Adv. Chem. Phys. 69 (1987) 1--62.
\newblock \href {http://dx.doi.org/10.1002/9780470142943.ch1}
  {\path{doi:10.1002/9780470142943.ch1}}.
\newline\urlprefix\url{http://dx.doi.org/10.1002/9780470142943.ch1}

\bibitem{Shepard1987}
R.~Shepard, \href{http://dx.doi.org/10.1002/9780470142943.ch2}{The
  multiconfiguration self-consistent field method}, Adv. Chem. Phys. 69 (2007)
  63--200.
\newblock \href {http://dx.doi.org/10.1002/9780470142943.ch2}
  {\path{doi:10.1002/9780470142943.ch2}}.
\newline\urlprefix\url{http://dx.doi.org/10.1002/9780470142943.ch2}

\bibitem{Nakano2000}
H.~Nakano, K.~Hirao,
  \href{http://www.sciencedirect.com/science/article/pii/S0009261499013640}{A
  quasi-complete active space self-consistent field method}, Chem. Phys. Lett.
  317~(1–2) (2000) 90 -- 96.
\newblock \href
  {http://dx.doi.org/http://dx.doi.org/10.1016/S0009-2614(99)01364-0}
  {\path{doi:http://dx.doi.org/10.1016/S0009-2614(99)01364-0}}.
\newline\urlprefix\url{http://www.sciencedirect.com/science/article/pii/S0009261499013640}

\bibitem{Tenno1996}
S.~Ten-no, S.~Iwata,
  \href{http://scitation.aip.org/content/aip/journal/jcp/105/9/10.1063/1.472231}{Multiconfiguration
  self-consistent field procedure employing linear combination of
  atomic-electron distributions}, J. Chem. Phys. 105~(9) (1996) 3604--3611.
\newblock \href {http://dx.doi.org/http://dx.doi.org/10.1063/1.472231}
  {\path{doi:http://dx.doi.org/10.1063/1.472231}}.
\newline\urlprefix\url{http://scitation.aip.org/content/aip/journal/jcp/105/9/10.1063/1.472231}

\bibitem{Gyoerffy2013}
W.~Gy\H{o}rffy, T.~Shiozaki, G.~Knizia, H.-J. Werner,
  \href{http://scitation.aip.org/content/aip/journal/jcp/138/10/10.1063/1.4793737}{Analytical
  energy gradients for second-order multireference perturbation theory using
  density fitting}, J. Chem. Phys. 138~(10) (2013) 104104.
\newblock \href {http://dx.doi.org/http://dx.doi.org/10.1063/1.4793737}
  {\path{doi:http://dx.doi.org/10.1063/1.4793737}}.
\newline\urlprefix\url{http://scitation.aip.org/content/aip/journal/jcp/138/10/10.1063/1.4793737}

\bibitem{Lindh2008}
F.~Aquilante, T.~B. Pedersen, R.~Lindh, B.~O. Roos, A.~S\'anchez~de Mer\'as,
  H.~Koch,
  \href{http://scitation.aip.org/content/aip/journal/jcp/129/2/10.1063/1.2953696}{Accurate
  ab initio density fitting for multiconfigurational self-consistent field
  methods}, J. Chem. Phys. 129~(2) (2008) 024113.
\newblock \href {http://dx.doi.org/http://dx.doi.org/10.1063/1.2953696}
  {\path{doi:http://dx.doi.org/10.1063/1.2953696}}.
\newline\urlprefix\url{http://scitation.aip.org/content/aip/journal/jcp/129/2/10.1063/1.2953696}

\bibitem{Hohenstein2015a}
E.~G. Hohenstein, N.~Luehr, I.~S. Ufimtsev, T.~J. Mart\'inez,
  \href{http://scitation.aip.org/content/aip/journal/jcp/142/22/10.1063/1.4921956}{An
  atomic orbital-based formulation of the complete active space self-consistent
  field method on graphical processing units}, J. Chem. Phys. 142~(22) (2015)
  224103.
\newblock \href {http://dx.doi.org/http://dx.doi.org/10.1063/1.4921956}
  {\path{doi:http://dx.doi.org/10.1063/1.4921956}}.
\newline\urlprefix\url{http://scitation.aip.org/content/aip/journal/jcp/142/22/10.1063/1.4921956}

\bibitem{Kim2015}
I.~Kim, S.~M. Parker, T.~Shiozaki,
  \href{http://dx.doi.org/10.1021/acs.jctc.5b00429}{Orbital optimization in the
  active space decomposition model}, J. Chem. Theory Comput. 11~(8) (2015)
  3636--3642.
\newblock \href
  {http://arxiv.org/abs/http://dx.doi.org/10.1021/acs.jctc.5b00429}
  {\path{arXiv:http://dx.doi.org/10.1021/acs.jctc.5b00429}}, \href
  {http://dx.doi.org/10.1021/acs.jctc.5b00429}
  {\path{doi:10.1021/acs.jctc.5b00429}}.
\newline\urlprefix\url{http://dx.doi.org/10.1021/acs.jctc.5b00429}

\bibitem{Levine}
B.~S. Fales, B.~G. Levine,
  \href{http://dx.doi.org/10.1021/acs.jctc.5b00634}{Nanoscale multireference
  quantum chemistry: Full configuration interaction on graphical processing
  units}, J. Chem. Theory Comput. 11~(10) (2015) 4708--4716.
\newblock \href
  {http://arxiv.org/abs/http://dx.doi.org/10.1021/acs.jctc.5b00634}
  {\path{arXiv:http://dx.doi.org/10.1021/acs.jctc.5b00634}}, \href
  {http://dx.doi.org/10.1021/acs.jctc.5b00634}
  {\path{doi:10.1021/acs.jctc.5b00634}}.
\newline\urlprefix\url{http://dx.doi.org/10.1021/acs.jctc.5b00634}

\bibitem{Chan2011b}
G.~K.-L. Chan, S.~Sharma,
  \href{http://dx.doi.org/10.1146/annurev-physchem-032210-103338}{The density
  matrix renormalization group in quantum chemistry}, Annu. Rev. Phys. Chem.
  62~(1) (2011) 465--481.
\newblock \href
  {http://arxiv.org/abs/http://dx.doi.org/10.1146/annurev-physchem-032210-103338}
  {\path{arXiv:http://dx.doi.org/10.1146/annurev-physchem-032210-103338}},
  \href {http://dx.doi.org/10.1146/annurev-physchem-032210-103338}
  {\path{doi:10.1146/annurev-physchem-032210-103338}}.
\newline\urlprefix\url{http://dx.doi.org/10.1146/annurev-physchem-032210-103338}

\bibitem{Sharma2014}
S.~Sharma, T.~Yanai, G.~H. Booth, C.~J. Umrigar, G.~K.-L. Chan,
  \href{http://scitation.aip.org/content/aip/journal/jcp/140/10/10.1063/1.4867383}{Spectroscopic
  accuracy directly from quantum chemistry: Application to ground and excited
  states of beryllium dimer}, J. Chem. Phys. 140~(10) (2014) 104112.
\newblock \href {http://dx.doi.org/http://dx.doi.org/10.1063/1.4867383}
  {\path{doi:http://dx.doi.org/10.1063/1.4867383}}.
\newline\urlprefix\url{http://scitation.aip.org/content/aip/journal/jcp/140/10/10.1063/1.4867383}

\bibitem{Booth2009}
G.~H. Booth, A.~J.~W. Thom, A.~Alavi,
  \href{http://scitation.aip.org/content/aip/journal/jcp/131/5/10.1063/1.3193710}{Fermion
  monte carlo without fixed nodes: A game of life, death, and annihilation in
  slater determinant space}, J. Chem. Phys. 131~(5) (2009) 054106.
\newblock \href {http://dx.doi.org/http://dx.doi.org/10.1063/1.3193710}
  {\path{doi:http://dx.doi.org/10.1063/1.3193710}}.
\newline\urlprefix\url{http://scitation.aip.org/content/aip/journal/jcp/131/5/10.1063/1.3193710}

\bibitem{Booth2012a}
G.~H. Booth, D.~Cleland, A.~Alavi, D.~P. Tew,
  \href{http://scitation.aip.org/content/aip/journal/jcp/137/16/10.1063/1.4762445}{An
  explicitly correlated approach to basis set incompleteness in full
  configuration interaction quantum monte carlo}, J. Chem. Phys. 137~(16)
  (2012) 164112.
\newblock \href {http://dx.doi.org/http://dx.doi.org/10.1063/1.4762445}
  {\path{doi:http://dx.doi.org/10.1063/1.4762445}}.
\newline\urlprefix\url{http://scitation.aip.org/content/aip/journal/jcp/137/16/10.1063/1.4762445}

\bibitem{Vogiatzis2015}
K.~D. Vogiatzis, G.~Li~Manni, S.~J. Stoneburner, D.~Ma, L.~Gagliardi,
  \href{http://dx.doi.org/10.1021/acs.jctc.5b00191}{Systematic expansion of
  active spaces beyond the casscf limit: A gasscf/splitgas benchmark study}, J.
  Chem. Theory Comput. 11~(7) (2015) 3010--3021.
\newblock \href
  {http://arxiv.org/abs/http://dx.doi.org/10.1021/acs.jctc.5b00191}
  {\path{arXiv:http://dx.doi.org/10.1021/acs.jctc.5b00191}}, \href
  {http://dx.doi.org/10.1021/acs.jctc.5b00191}
  {\path{doi:10.1021/acs.jctc.5b00191}}.
\newline\urlprefix\url{http://dx.doi.org/10.1021/acs.jctc.5b00191}

\bibitem{DePrince2016}
J.~Fosso-Tande, T.-S. Nguyen, G.~Gidofalvi, A.~E. DePrince,
  \href{http://dx.doi.org/10.1021/acs.jctc.6b00190}{Large-scale variational
  two-electron reduced-density-matrix-driven complete active space
  self-consistent field methods}, J. Chem. Theory Comput. 12~(5) (2016)
  2260--2271.
\newblock \href
  {http://arxiv.org/abs/http://dx.doi.org/10.1021/acs.jctc.6b00190}
  {\path{arXiv:http://dx.doi.org/10.1021/acs.jctc.6b00190}}, \href
  {http://dx.doi.org/10.1021/acs.jctc.6b00190}
  {\path{doi:10.1021/acs.jctc.6b00190}}.
\newline\urlprefix\url{http://dx.doi.org/10.1021/acs.jctc.6b00190}

\bibitem{Ghosh2008}
D.~Ghosh, J.~Hachmann, T.~Yanai, G.~K.-L. Chan,
  \href{http://scitation.aip.org/content/aip/journal/jcp/128/14/10.1063/1.2883976;jsessionid=2aMke8TmEjipekjqxpWZOQx0.x-aip-live-03}{Orbital
  optimization in the density matrix renormalization group, with applications
  to polyenes and β-carotene}, J. Chem. Phys. 128~(14).
\newblock \href {http://dx.doi.org/http://dx.doi.org/10.1063/1.2883976}
  {\path{doi:http://dx.doi.org/10.1063/1.2883976}}.
\newline\urlprefix\url{http://scitation.aip.org/content/aip/journal/jcp/128/14/10.1063/1.2883976;jsessionid=2aMke8TmEjipekjqxpWZOQx0.x-aip-live-03}

\bibitem{Zgid2008}
D.~Zgid, M.~Nooijen,
  \href{http://scitation.aip.org/content/aip/journal/jcp/128/14/10.1063/1.2883981}{The
  density matrix renormalization group self-consistent field method: Orbital
  optimization with the density matrix renormalization group method in the
  active space}, J. Chem. Phys. 128~(14).
\newblock \href {http://dx.doi.org/http://dx.doi.org/10.1063/1.2883981}
  {\path{doi:http://dx.doi.org/10.1063/1.2883981}}.
\newline\urlprefix\url{http://scitation.aip.org/content/aip/journal/jcp/128/14/10.1063/1.2883981}

\bibitem{Yanai2009}
T.~Yanai, Y.~Kurashige, D.~Ghosh, G.~K.-L. Chan,
  \href{http://dx.doi.org/10.1002/qua.22099}{Accelerating convergence in
  iterative solution for large-scale complete active space
  self-consistent-field calculations}, Int. J. Quant. Chem. 109~(10) (2009)
  2178--2190.
\newblock \href {http://dx.doi.org/10.1002/qua.22099}
  {\path{doi:10.1002/qua.22099}}.
\newline\urlprefix\url{http://dx.doi.org/10.1002/qua.22099}

\bibitem{Ma2013}
Y.~Ma, H.~Ma,
  \href{http://scitation.aip.org/content/aip/journal/jcp/138/22/10.1063/1.4809682}{Assessment
  of various natural orbitals as the basis of large active space density-matrix
  renormalization group calculations}, J. Chem. Phys. 138~(22).
\newblock \href {http://dx.doi.org/http://dx.doi.org/10.1063/1.4809682}
  {\path{doi:http://dx.doi.org/10.1063/1.4809682}}.
\newline\urlprefix\url{http://scitation.aip.org/content/aip/journal/jcp/138/22/10.1063/1.4809682}

\bibitem{Wouters2014}
S.~Wouters, T.~Bogaerts, P.~Van Der~Voort, V.~Van~Speybroeck, D.~Van~Neck,
  \href{http://scitation.aip.org/content/aip/journal/jcp/140/24/10.1063/1.4885815}{Communication:
  Dmrg-scf study of the singlet, triplet, and quintet states of oxo-mn(salen)},
  J. Chem. Phys. 140~(24).
\newblock \href {http://dx.doi.org/http://dx.doi.org/10.1063/1.4885815}
  {\path{doi:http://dx.doi.org/10.1063/1.4885815}}.
\newline\urlprefix\url{http://scitation.aip.org/content/aip/journal/jcp/140/24/10.1063/1.4885815}

\bibitem{fciqmccasscf}
R.~E. Thomas, Q.~Sun, A.~Alavi, G.~H. Booth,
  \href{http://dx.doi.org/10.1021/acs.jctc.5b00917}{Stochastic
  multiconfigurational self-consistent field theory}, J. Chem. Theory Comput.
  11~(11) (2015) 5316--5325.
\newblock \href
  {http://arxiv.org/abs/http://dx.doi.org/10.1021/acs.jctc.5b00917}
  {\path{arXiv:http://dx.doi.org/10.1021/acs.jctc.5b00917}}, \href
  {http://dx.doi.org/10.1021/acs.jctc.5b00917}
  {\path{doi:10.1021/acs.jctc.5b00917}}.
\newline\urlprefix\url{http://dx.doi.org/10.1021/acs.jctc.5b00917}

\bibitem{Alavi2016}
G.~Li~Manni, S.~D. Smart, A.~Alavi,
  \href{http://dx.doi.org/10.1021/acs.jctc.5b01190}{Combining the complete
  active space self-consistent field method and the full configuration
  interaction quantum monte carlo within a super-ci framework, with application
  to challenging metal-porphyrins}, J. Chem. Theory Comput. 12~(3) (2016)
  1245--1258.
\newblock \href
  {http://arxiv.org/abs/http://dx.doi.org/10.1021/acs.jctc.5b01190}
  {\path{arXiv:http://dx.doi.org/10.1021/acs.jctc.5b01190}}, \href
  {http://dx.doi.org/10.1021/acs.jctc.5b01190}
  {\path{doi:10.1021/acs.jctc.5b01190}}.
\newline\urlprefix\url{http://dx.doi.org/10.1021/acs.jctc.5b01190}

\bibitem{Reiher2016}
Y.~Ma, S.~Knecht, S.~Keller, M.~Reiher, Second-order self-consistent-field
  density-matrix renormalization group, arXiv:1611.05972 [physics.chem-ph]\href
  {http://arxiv.org/abs/1611.05972} {\path{arXiv:1611.05972}}.

\bibitem{PYSCF}
Q.~Sun, Python module for quantum chemistry program,
  \url{https://github.com/sunqm/pyscf.git} (2014).

\bibitem{Davidson1975}
E.~R. Davidson,
  \href{http://www.sciencedirect.com/science/article/pii/0021999175900650}{The
  iterative calculation of a few of the lowest eigenvalues and corresponding
  eigenvectors of large real-symmetric matrices}, J. Comput. Phys. 17~(1)
  (1975) 87 -- 94.
\newblock \href {http://dx.doi.org/10.1016/0021-9991(75)90065-0}
  {\path{doi:10.1016/0021-9991(75)90065-0}}.
\newline\urlprefix\url{http://www.sciencedirect.com/science/article/pii/0021999175900650}

\bibitem{Sharma2012}
S.~Sharma, G.~K.-L. Chan,
  \href{http://scitation.aip.org/content/aip/journal/jcp/136/12/10.1063/1.3695642}{Spin-adapted
  density matrix renormalization group algorithms for quantum chemistry}, J.
  Chem. Phys. 136~(12).
\newblock \href {http://dx.doi.org/http://dx.doi.org/10.1063/1.3695642}
  {\path{doi:http://dx.doi.org/10.1063/1.3695642}}.
\newline\urlprefix\url{http://scitation.aip.org/content/aip/journal/jcp/136/12/10.1063/1.3695642}

\bibitem{Knizia2012b}
G.~Knizia, G.~K.-L. Chan,
  \href{http://link.aps.org/doi/10.1103/PhysRevLett.109.186404}{Density matrix
  embedding: A simple alternative to dynamical mean-field theory}, Phys. Rev.
  Lett. 109 (2012) 186404.
\newblock \href {http://dx.doi.org/10.1103/PhysRevLett.109.186404}
  {\path{doi:10.1103/PhysRevLett.109.186404}}.
\newline\urlprefix\url{http://link.aps.org/doi/10.1103/PhysRevLett.109.186404}

\bibitem{QimingSun2014}
Q.~Sun, G.~K.-L. Chan, \href{http://dx.doi.org/10.1021/ct500512f}{Exact and
  optimal quantum mechanics/molecular mechanics boundaries}, J. Chem. Theory
  Comput. 10~(9) (2014) 3784--3790.
\newblock \href {http://arxiv.org/abs/http://dx.doi.org/10.1021/ct500512f}
  {\path{arXiv:http://dx.doi.org/10.1021/ct500512f}}, \href
  {http://dx.doi.org/10.1021/ct500512f} {\path{doi:10.1021/ct500512f}}.
\newline\urlprefix\url{http://dx.doi.org/10.1021/ct500512f}

\bibitem{practicalDMET}
S.~Wouters, C.~A. Jim\'enez-Hoyos, Q.~Sun, G.~K.-L. Chan,
  \href{http://dx.doi.org/10.1021/acs.jctc.6b00316}{A practical guide to
  density matrix embedding theory in quantum chemistry}, J. Chem. Theory
  Comput. 12~(6) (2016) 2706--2719.
\newblock \href
  {http://arxiv.org/abs/http://dx.doi.org/10.1021/acs.jctc.6b00316}
  {\path{arXiv:http://dx.doi.org/10.1021/acs.jctc.6b00316}}, \href
  {http://dx.doi.org/10.1021/acs.jctc.6b00316}
  {\path{doi:10.1021/acs.jctc.6b00316}}.
\newline\urlprefix\url{http://dx.doi.org/10.1021/acs.jctc.6b00316}

\bibitem{Sayfutyarova2017}
E.~R. Sayfutyarova, G.~Knizia, Q.~Sun, G.~K.-L. Chan, Automatic construction of
  molecular active spaces from atomic valence orbitalsIn preparation.

\bibitem{Edwards1986}
W.~D. Edwards, B.~Weiner, M.~C. Zerner,
  \href{http://dx.doi.org/10.1021/ja00269a012}{On the low-lying states and
  electronic spectroscopy of iron(ii) porphine}, J. Am. Chem. Soc. 108~(9)
  (1986) 2196--2204.
\newblock \href {http://arxiv.org/abs/http://dx.doi.org/10.1021/ja00269a012}
  {\path{arXiv:http://dx.doi.org/10.1021/ja00269a012}}, \href
  {http://dx.doi.org/10.1021/ja00269a012} {\path{doi:10.1021/ja00269a012}}.
\newline\urlprefix\url{http://dx.doi.org/10.1021/ja00269a012}

\bibitem{Rovira1997}
C.~Rovira, K.~Kunc, J.~Hutter, P.~Ballone, M.~Parrinello,
  \href{http://dx.doi.org/10.1021/jp9722115}{Equilibrium geometries and
  electronic structure of iron-porphyrin complexes:  a density functional
  study}, J. Phys. Chem. A 101~(47) (1997) 8914--8925.
\newblock \href {http://arxiv.org/abs/http://dx.doi.org/10.1021/jp9722115}
  {\path{arXiv:http://dx.doi.org/10.1021/jp9722115}}, \href
  {http://dx.doi.org/10.1021/jp9722115} {\path{doi:10.1021/jp9722115}}.
\newline\urlprefix\url{http://dx.doi.org/10.1021/jp9722115}

\bibitem{Choe1999}
Y.-K. Choe, T.~Nakajima, K.~Hirao, R.~Lindh,
  \href{http://scitation.aip.org/content/aip/journal/jcp/111/9/10.1063/1.479687}{Theoretical
  study of the electronic ground state of iron(ii) porphine. ii}, J. Chem.
  Phys. 111~(9) (1999) 3837--3845.
\newblock \href {http://dx.doi.org/http://dx.doi.org/10.1063/1.479687}
  {\path{doi:http://dx.doi.org/10.1063/1.479687}}.
\newline\urlprefix\url{http://scitation.aip.org/content/aip/journal/jcp/111/9/10.1063/1.479687}

\bibitem{Oliveira2001}
K.~Oliveira, M.~Trsic,
  \href{http://www.sciencedirect.com/science/article/pii/S0166128000007788}{Comparative
  theoretical study of the electronic structures and electronic spectra of
  fe2+-, fe+3-porphyrin and free base porphyrin}, J. Mol. Struct. (THEOCHEM)
  539~(1–3) (2001) 107 -- 117.
\newblock \href
  {http://dx.doi.org/http://dx.doi.org/10.1016/S0166-1280(00)00778-8}
  {\path{doi:http://dx.doi.org/10.1016/S0166-1280(00)00778-8}}.
\newline\urlprefix\url{http://www.sciencedirect.com/science/article/pii/S0166128000007788}

\bibitem{Pierloot2003}
K.~Pierloot,
  \href{http://www.tandfonline.com/doi/abs/10.1080/0026897031000109356}{The
  caspt2 method in inorganic electronic spectroscopy: from ionic transition
  metal to covalent actinide complexes∗}, Mol. Phys. 101~(13) (2003)
  2083--2094.
\newblock \href
  {http://arxiv.org/abs/http://www.tandfonline.com/doi/pdf/10.1080/0026897031000109356}
  {\path{arXiv:http://www.tandfonline.com/doi/pdf/10.1080/0026897031000109356}},
  \href {http://dx.doi.org/10.1080/0026897031000109356}
  {\path{doi:10.1080/0026897031000109356}}.
\newline\urlprefix\url{http://www.tandfonline.com/doi/abs/10.1080/0026897031000109356}

\bibitem{Groenhof2005}
A.~R. Groenhof, M.~Swart, A.~W. Ehlers, K.~Lammertsma,
  \href{http://dx.doi.org/10.1021/jp0441442}{Electronic ground states of iron
  porphyrin and of the first species in the catalytic reaction cycle of
  cytochrome p450s}, J. Phys. Chem. A. 109~(15) (2005) 3411--3417.
\newblock \href {http://arxiv.org/abs/http://dx.doi.org/10.1021/jp0441442}
  {\path{arXiv:http://dx.doi.org/10.1021/jp0441442}}, \href
  {http://dx.doi.org/10.1021/jp0441442} {\path{doi:10.1021/jp0441442}}.
\newline\urlprefix\url{http://dx.doi.org/10.1021/jp0441442}

\bibitem{Liao2006}
M.-S. Liao, J.~D. Watts, M.-J. Huang,
  \href{http://dx.doi.org/10.1002/jcc.20458}{Assessment of the performance of
  density-functional methods for calculations on iron porphyrins and related
  compounds}, J. Comput. Chem. 27~(13) (2006) 1577--1592.
\newblock \href {http://dx.doi.org/10.1002/jcc.20458}
  {\path{doi:10.1002/jcc.20458}}.
\newline\urlprefix\url{http://dx.doi.org/10.1002/jcc.20458}

\bibitem{Shaik2011}
H.~Chen, W.~Lai, S.~Shaik,
  \href{http://dx.doi.org/10.1021/jp110016u}{Multireference and
  multiconfiguration ab initio methods in heme-related systems: What have we
  learned so far?}, J. Chem. Phys. 115~(8) (2011) 1727--1742.
\newblock \href {http://arxiv.org/abs/http://dx.doi.org/10.1021/jp110016u}
  {\path{arXiv:http://dx.doi.org/10.1021/jp110016u}}, \href
  {http://dx.doi.org/10.1021/jp110016u} {\path{doi:10.1021/jp110016u}}.
\newline\urlprefix\url{http://dx.doi.org/10.1021/jp110016u}

\bibitem{Swart2013}
M.~Swart, \href{http://dx.doi.org/10.1002/qua.24255}{Spin states of
  (bio)inorganic systems: Successes and pitfalls}, Int. J. Quant. Chem. 113~(1)
  (2013) 2--7.
\newblock \href {http://dx.doi.org/10.1002/qua.24255}
  {\path{doi:10.1002/qua.24255}}.
\newline\urlprefix\url{http://dx.doi.org/10.1002/qua.24255}

\bibitem{Olivares-Amaya2015}
R.~Olivares-Amaya, W.~Hu, N.~Nakatani, S.~Sharma, J.~Yang, G.~K.-L. Chan,
  \href{http://scitation.aip.org/content/aip/journal/jcp/142/3/10.1063/1.4905329}{The
  ab-initio density matrix renormalization group in practice}, J. Chem. Phys.
  142~(3) (2015) 034102.
\newblock \href {http://dx.doi.org/http://dx.doi.org/10.1063/1.4905329}
  {\path{doi:http://dx.doi.org/10.1063/1.4905329}}.
\newline\urlprefix\url{http://scitation.aip.org/content/aip/journal/jcp/142/3/10.1063/1.4905329}

\bibitem{Vancoillie2011}
S.~Vancoillie, H.~Zhao, V.~T. Tran, M.~F.~A. Hendrickx, K.~Pierloot,
  \href{http://dx.doi.org/10.1021/ct200597h}{Multiconfigurational second-order
  perturbation theory restricted active space (raspt2) studies on mononuclear
  first-row transition-metal systems}, J. Chem. Theory Comput. 7~(12) (2011)
  3961--3977.
\newblock \href {http://arxiv.org/abs/http://dx.doi.org/10.1021/ct200597h}
  {\path{arXiv:http://dx.doi.org/10.1021/ct200597h}}, \href
  {http://dx.doi.org/10.1021/ct200597h} {\path{doi:10.1021/ct200597h}}.
\newline\urlprefix\url{http://dx.doi.org/10.1021/ct200597h}

\end{thebibliography}

\clearpage

\begin{table}
  \centering
  \caption{Co-iterative algorithm for orbital optimization}
\begin{tabular}{ccclcccccc}
  \hline
  Davidson  &              & \multicolumn{2}{l}{AH matrix elements} && Davidson \\
  \cline{1-1} \cline{3-4} \cline{6-6} \\
  iteration & AH iteration & Hessian & Gradient && space size \\
  \hline
  0 &                      & $\mathcal{H}$ & $\mathcal{G}_{[0]}$                                                 && 1 \\
  1 &                      & $\mathcal{H}$ & $\mathcal{G}_{[0]}$                                                 && 2 \\
  2 & $\mathbf{R}^1_{[0]}$ & $\mathcal{H}$ & $\mathcal{G}_{[1]}=\mathcal{G}_{[0]}+\mathcal{H}\mathbf{R}^1_{[0]}$ && 3 \\
  3 & $\mathbf{R}^1_{[1]}$ & $\mathcal{H}$ & $\mathcal{G}_{[2]}=\mathcal{G}_{[1]}+\mathcal{H}\mathbf{R}^1_{[1]}$ && 4 \\
  4 & $\mathbf{R}^1_{[2]}$ & $\mathcal{H}$ & $\mathcal{G}_{[3]}=\mathcal{G}_{[2]}+\mathcal{H}\mathbf{R}^1_{[2]}$ && 5 \\
  5 & \vdots               & $\mathcal{H}$ & \vdots                                                              && 6 \\
  \hline
\end{tabular}
  \label{tab:fwah}
\end{table}

\begin{table}
  \centering
  \caption{The types of two-electron integrals required by different approximations for
  MO-driven and AO-driven algorithms.}
\begin{tabular}[htp]{llllll}
  \hline
       & MO-driven & AO-driven \\
  \hline
  FEP1 & $(AA|\!*\!*), (A\!*\!|A*)$,                   & $(AA|\!*\!*), (A\!*\!|A*)$ \\
       & $(CC|VV), (CV|CV), (CV|AC)$, & \\
       & $(CV|AV), (AV|CC), (AC|VV)$  & \\
  DEP1 & $(AA|\!*\!*), (A\!*\!|A*)$,                   & $(AA|\!*\!*), (A\!*\!|A*)$ \\
       & $(CC|C*), (CC|V*), (CV|C*), (CV|A*), (AV|C*)$,         & \\
       & $(CA|C*), (CA|V*)$                            & \\
  DEP2 & $(A\!*\!|\!*\!*), (CC|\!*\!*), (CV|\!*\!*)$   & $(A\!*\!|\!*\!*)$ \\
  DEP4 & $(*\!*\!|\!*\!*)$                             & $(*\!*\!|\!*\!*)$ \\
  \hline
\end{tabular}
  \label{tab:costs}
\end{table}

\begin{table}[htp]
  \centering
  \begin{tabular}{lllllll}
    \hline
    Notation & \multicolumn{1}{c}{Description} \\
    \hline
    FEP1   & Frozen $\mathcal{H}^{oo}$. First order $\mathbf{R}$
             expansion for $\mathcal{G}^o$ and CI Hamiltonian \\
    DEP1   & Frozen $\mathcal{H}^{oo}$. First order $\mathbf{T}$
             expansion for $\mathcal{G}^o$ and CI Hamiltonian \\
    DEP2   & Frozen $\mathcal{H}^{oo}$. Second order $\mathbf{T}$
             expansion for $\mathcal{G}^o$ and CI Hamiltonian \\
    DEP1+  & First order $\mathbf{T}$ expansion for
             $\mathcal{H}^{oo}$, $\mathcal{G}^o$ and CI Hamiltonian \\
    DEP2+  & Second order $\mathbf{T}$ expansion for
             $\mathcal{H}^{oo}$, $\mathcal{G}^o$ and CI Hamiltonian \\
    DEP4+  & Exact 
             $\mathcal{H}^{oo}$, $\mathcal{G}^o$ and CI Hamiltonian \\
    \hline
  \end{tabular}
  \caption{DEP/FEP approximations for microiteration updates.}
  \label{tab:approx1step}
\end{table}

\begin{table}
  \centering
  \caption{Number of macroiterations required for convergence using different DEP/FEP approximations. The convergence
    threshold is $10^{-8}$ Hartrees.}
  \begin{tabular}{llllllllllllll}
    \hline
           & N$_2$& CO   & HF  &C$_2$&C$_2$&O$_3$ &NO$_2$&CH$_2$ & HCHO & C$_6$H$_6$ \\
    CASSCF &(10,8)&(10,8)&(8,9)&(8,8)&(8,8)&(12,9)&(5,6) &(6,14) &(12,10)& (6,6)\\
    State  & $^1\Sigma_{g+}$ & $^1\Sigma_{g+}$ & $^1\Sigma_{g+}$ & $^1\Sigma_{g+}$ & $^3\Pi_{u+}$
                                                & $^1A_1$ & $^2A_1$ & $^3B_2$ & $^1A_1$ & $^1A_g$ \\
    \hline
    DEP4+  & 3    & 4    & 5   & 3   & 3   & 3    & 5    & 5     & 6    & 3 \\
    DEP2+  & 3    & 4    & 5   & 3   & 3   & 3    & 5    & 5     & 6    & 3 \\
    DEP1+  & 3    & 4    & 5   & 3   & 3   & 4    & 5    & 5     & 6    & 3 \\
    DEP2   & 3    & 4    & 5   & 3   & 3   & 3    & 5    & 5     & 6    & 3 \\
    DEP1   & 3    & 4    & 5   & 3   & 3   & 4    & 5    & 5     & 6    & 3 \\
    FEP1   & 3    & 4    & 5   & 3   & 3   & 4    & 5    & 7     & 6    & 3 \\
    2-step & 10   & 13   & 17  & 9   & 11  & 9    & 17   & 22    & 21   & 7 \\
    \hline
  \end{tabular}
  \label{tab:dep1test}
\end{table}

\begin{table}
  \centering
  \caption{Number of macro iterations, micro iterations and J,K calls to
  converge the singlet, triplet, and quintet states of Fe(\Roman{2})-porphine
  with CASSCF(8,11) and DMRG-CASSCF(22,27)}
  \begin{threeparttable}
  \begin{tabular}{lllllllll}
    \hline
    State & Active space & Basis          & Energy        & macro & micro & J,K calls \\
    \hline
    $^1A_{g}$ & CAS(8,11) & DZ & -2244.7656583 & 4     & 13    & 63    \\
              &          & TZ & -2244.9928018 & 5     & 15    & 74    \\
              &          & QZ & -2245.0550513 & 5     & 15    & 72    \\
    $^3B_{1g}$& CAS(8,11) 
                         & DZ & -2244.8155591 & 18    & 54    & 274   \\
              &          & TZ & -2245.0429961 & 10    & 34    & 201   \\
              &          & QZ & -2245.1050841 & 9     & 33    & 197   \\
              &          & 5Z & -2245.1187001 & 9     & 32    & 203   \\
    $^3B_{3g}$&CAS(8,11)& DZ & -2244.8113231 & 5     & 15    & 74    \\
              &         & TZ & -2245.0378914 & 6     & 18    & 93    \\
              &         & QZ & -2245.0999279 & 6     & 18    & 100   \\
    $^5A_{g}$&CAS(8,11) & DZ & -2244.8291051 & 5     & 17    & 88    \\
             &          & TZ & -2245.0559164 & 6     & 18    & 91    \\
             &          & QZ & -2245.1180998 & 6     & 21    & 126   \\
             &          & 5Z & -2245.1316309 & 6     & 19    & 92    \\
    $^5B_{2g}$&CAS(8,11)& DZ & -2244.8204477 & 7     & 22    & 113   \\
              &         & TZ & -2245.0474655 & 5     & 15    & 76    \\
              &         & QZ & -2245.1095986 & 6     & 19    & 102   \\
    $^3B_{1g}$&CAS(22,27)& DZ & -2245.0006085 & 8     & 26    & 136   \\
              &          & QZ & -2245.2917119 & 7     & 22    & 121   \\
              &          & 5Z & -2245.3060865 & 8     & 20    & 94    \\
    $^3B_{3g}$&CAS(22,27)& DZ& -2244.9974501 & 8     & 29    & 145   \\
    $^5A_{g}$&CAS(22,27)& DZ & -2245.0062936 & 8     & 28    & 148   \\
             &          & QZ & -2245.2974638 & 10    & 33    & 198   \\
             &          & 5Z & -2245.311861  & 8     & 29    & 149   \\
    $^5B_{2g}$&CAS(22,27)& DZ& -2244.9985422 & 8     & 29    & 150   \\
    \hline
  \end{tabular}
  \end{threeparttable}
  \label{tab:feporcas11}
\end{table}

\clearpage

\end{document}